# GROW: A Conversational AI Coach for Goals, Reflection, Optimism, and Well-Being

Keya Shah
kts5726@nyu.edu
New York University
Abu Dhabi, United Arab Emirates

Himanshi Lalwani
hl3937@nyu.edu
New York University
Abu Dhabi, United Arab Emirates

Hanan Salam
hs4461@nyu.edu
New York University
Abu Dhabi, United Arab Emirates

## Abstract

College students face well-being challenges driven by academic pressure, financial strain, and social expectations. While campus counseling and student-success programs offer support, access is often limited by stigma, waitlists, and scheduling constraints. Existing digital tools focus on emotional check-ins or chatbots and may overlook structured goal setting and aligning goals with personal values. We present *GROW*, a goal-centered well-being coaching system that puts values-aligned goals at the center of the student experience. *GROW* combines the SMART framework with principles from Acceptance and Commitment Therapy in a conversational AI coach that helps students clarify aspirations, break them into concrete steps, and reflect on progress. The system links action plans with Google Calendar, sends reminders, and provides a dashboard that shows progress and engagement. We evaluated *GROW* through interviews with clinical psychologists, student-success staff, and faculty, followed by a one-week deployment with 30 undergraduates. Findings offer design implications for interactive systems that support engagement, accountability, and sense of purpose in higher education.

## 1 Introduction

University life represents a formative yet demanding period marked by academic pressure, financial strain, social expectations, and identity development. These overlapping stressors place students' psychological well-being under constant strain [53]. Recent evidence indicates that mental health concerns are pervasive in this population, with depression and anxiety rates among university-aged individuals remaining alarmingly high [1]. Some estimates suggest that in certain samples, a majority of young adults meet diagnostic thresholds for these disorders during emerging adulthood [53]. Even among those who do not meet clinical criteria, many report feeling emotionally exhausted, socially disconnected, and uncertain about their sense of purpose, often as a result of academic pressure and life transitions [65, 66, 74]. Such experiences hinder motivation [81], academic performance [32], and life satisfaction [26].

In response, universities have introduced a range of well-being initiatives such as counseling centers and peer mentoring programs [49]. University counseling is one of the most established forms of support, and short-term interventions have been shown to improve mental health and academic outcomes [20, 55, 83]. However, despite these demonstrated benefits, counseling services are often stretched beyond capacity, resulting in long waitlists, inconsistent care, and limited accessibility for those in need [53]. Many students also face practical barriers such as scheduling conflicts or discomfort seeking help due to concerns about stigma and privacy [63]. To complement formal services, institutions have adopted peer support and mentoring initiatives that bring students together to share experiences, offer emotional support, and reduce feelings of isolation [75]. The informal and relatable nature of such programs can lower barriers to help-seeking and encourage students to reach out earlier [16]. Yet, ensuring that peer supporters receive adequate training, supervision, and boundaries remains a challenge, particularly when responding to students in crisis [9].

These combined structural and psychological barriers have driven growing interest in digital well-being interventions [14, 18, 79]. Such tools offer flexible, private, and scalable support that can complement traditional counseling while fitting more easily into students' daily routines [43]. For instance, digital interventions have been found to reduce symptoms of anxiety and depression [43, 53], and their affordability, anonymity, and on-demand access make them particularly appealing for students hesitant to engage in face-to-face therapy [11, 70]. However, most existing systems [23, 25] are grounded in cognitive-behavioral therapy (CBT) principles [31] that emphasize symptom management through self-monitoring or cognitive reframing [33]. While these systems can alleviate distress, they focus primarily on reducing negative emotions rather than cultivating the motivation, sense of purpose, and reflective habits that support long-term flourishing. In practice, this has led to fragmented designs: apps that offer journaling, meditation, or mood tracking without integrating these practices into students' daily workflows or connecting reflection to concrete actions in their academic and personal lives [24, 35, 85, 86]. Moreover, prior analyses of mental health apps suggest that sustaining engagement over time is challenging in many unguided systems, with substantial early drop-off observed across tools and techniques [5].

Against this backdrop, emerging AI technologies, particularly conversational agents, offer new possibilities for more interactive, personalized mental health and well-being support, including chatbots and coaching systems that can extend access and tailor support to students' needs [3, 44, 48, 77]. At the same time, recent reviews highlight substantial ethical and design challenges around autonomy, dependence, privacy, safety, and the appropriate role of AI relative to human care [19, 56, 72]. Within HCI and DIS, prior work on digital mental health and reflection has articulated design resources for reflection-supporting technologies [7, 45] and user-centered digital interventions for university students [62], yet comparatively little research has examined goal-oriented AI systems that embed values-based reflection into students' everyday academic tools while explicitly maintaining non-clinical boundaries [3, 77].

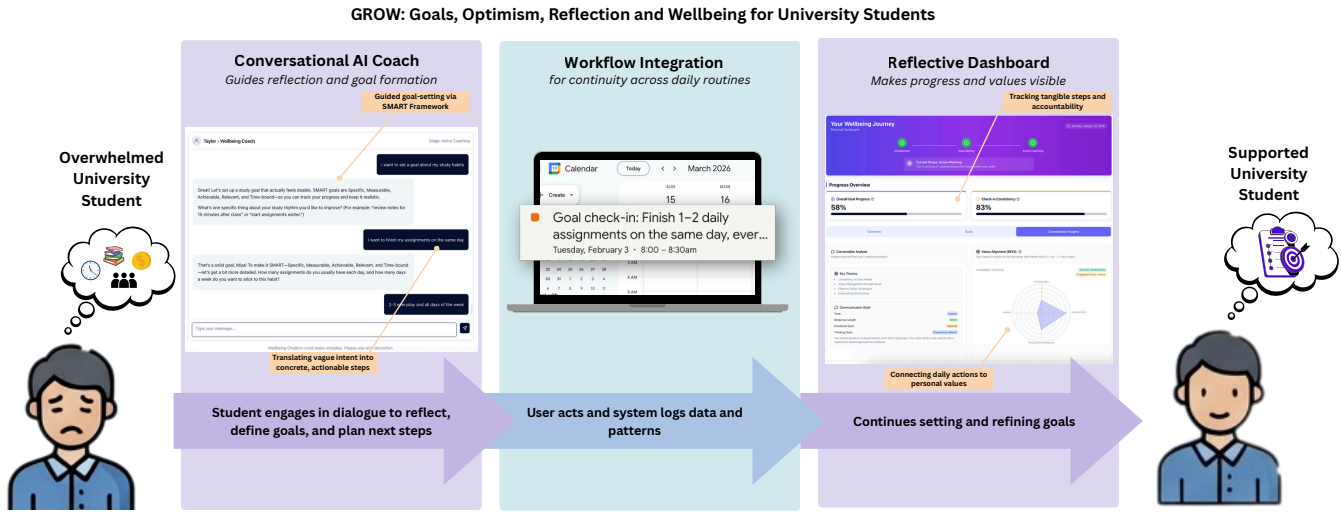

Figure 1: *GROW* System Overview. The figure illustrates how GROW supports students across a cycle of goal articulation, follow-through, and reflection. A student engages with the conversational AI coach when feeling overwhelmed by competing demands or to set goals. Through dialogue, the coach supports reflection, clarifies priorities, and guides goal formation using structured, non-clinical interaction strategies, including SMART goal setting and values-based prompts. Goals and commitments expressed in conversation are then connected to students' existing routines through workflow integrations, such as scheduled check-ins and calendar reminders. These ongoing interactions are surfaced in a reflective dashboard, where students can review active and completed goals, monitor progress and consistency, and explore aggregated conversation insights, including key themes, communication patterns, and values alignment. Together, these components aim to support continued goal refinement and re-engagement over time, framing goal work as an ongoing, adaptable process embedded within students' everyday lives.

To address these gaps, we explore how an AI system might support student well-being through structured goal setting and values-based reflection. Goal-setting approaches such as the SMART Goals framework provide a concrete scaffold for turning intentions into actionable plans [21], while Acceptance and Commitment Therapy (ACT) emphasizes values clarification and committed action as pathways to psychological flexibility [41]. We treat these frameworks as design resources for building non-clinical, everyday support rather than as a replacement for professional care. Drawing on these frameworks, we present *GROW*, a goal-oriented AI well-being system for university students that combines (1) a conversational coach to support SMART goal setting and lightweight values-based reflection, (2) a dashboard that visualizes progress and reflections over time, and (3) calendar and email integrations that situate goal work in students' existing academic workflows. We ask:

**RQ: How do university students experience and use a goal-oriented AI well-being system designed around structured goal setting and values-based reflection?**

To investigate this question, we conducted a two-part evaluation: a formative study with practitioners (e.g., clinical psychologists and student success coaches) to elicit design constraints, opportunities, and risks, followed by a one-week deployment with 30 university students to examine real-world use, breakdowns, and perceived value in daily life. This paper contributes:

- *GROW*, an AI-enabled well-being system that integrates structured goal setting, values-based reflection, and support across a conversational coach, progress dashboard, and calendar- and email-based reminders for university students.
- **Empirical findings** from expert feedback and an in-situ student deployment describing how students used the system, what they found supportive, and where the design introduced friction or disengagement.
- **Design implications** for goal-oriented AI well-being systems that aim to support action and autonomy while maintaining non-clinical boundaries.

## 2 Related Work

### 2.1 Digital Well-being Systems for Students

Digital tools for mental health and well-being have expanded rapidly in recent years, particularly conversational systems designed to provide accessible, low-stigma support [24, 27, 35, 60, 69, 85]. Many of these systems draw on cognitive behavioral therapy (CBT) principles [23, 25], delivering structured techniques such as cognitive reframing, journaling, or activity scheduling through dialogue. For example, Woebot [24] demonstrated that brief, daily CBT-informed check-ins could reduce short-term depressive symptoms among college students. MindfulDiary [35] supported reflective journaling through conversational prompts and summaries, while CRBot [85] used Socratic questioning to guide cognitive restructuring in supervised settings. Wayhaven [69] adapted conversational support for university students by integrating campus resources and micro-interventions into chat-based interactions. More recent LLM-powered systems, including YouMatter [60], SouLLMate [27], and a WHO-endorsed behavioral activation chatbot [68], further demonstrate the feasibility of scalable conversational tools that offer personalized emotional support, mood tracking, and activity planning, including for student populations.

Across this body of work, conversational systems have shown in some studies to support reflection, emotional expression, and short-term coping. However, many systems tend to focus on momentary relief or symptom management rather than sustained engagement with personally meaningful goals and follow-through over time. Reflective practices such as journaling or mood check-ins are often decoupled from mechanisms that help users translate insight into concrete action within the context of their ongoing academic and personal routines. From a design perspective, this points to a limitation in how conversational support is typically designed: most systems emphasize episodic interactions within a standalone app, rather than continuity, longitudinal follow-through, and integration with students' existing academic tools and workflows. Our system, GROW addresses this gap by placing values-aligned goal setting at the center of interaction and treating goals as evolving commitments rather than one-time configurations. It explicitly links reflective dialogue to concrete actions, reminders, and progress tracking over time, and situates this support within the academic tools students already use.

## 2.2 ACT-Based Interventions for Students

Acceptance and Commitment Therapy (ACT) [41] offers a values-oriented approach to well-being that emphasizes psychological flexibility, or the ability to act in accordance with one's values despite difficult thoughts or emotions. ACT is commonly described through six interrelated processes: acceptance, cognitive defusion, present-moment awareness, self-as-context, values clarification, and committed action [29]. While ACT originates in clinical therapy, several of its components, particularly values reflection and committed action, have been adapted for non-clinical and digital interventions [41, 46]. A recent scoping review identifies dozens of ACT-based interventions in higher education, across both in-person and online formats [17].

Within ACT-based interventions, the Bull's-Eye Values Survey (BEVS) [52] is a well-established tool for operationalizing values reflection. BEVS is a visual instrument that helps people assess how closely their actions align with personally important values across domains such as education, relationships, health, and leisure. It has been validated with university students and clinical populations, demonstrating reliability and sensitivity to change in ACT-based interventions [8, 52]. Its brief, structured format and domain-based framing make it amenable to repeated use and adaptation within digital systems.

Several web-based ACT programs have been developed for university and college students, including guided and unguided interventions, self-help courses, and blended online formats [34, 38, 42, 46, 67, 82, 84]. Collectively, these studies show that ACT processes can be delivered online to students and can improve a range of mental health and academic outcomes. At the same time, most of these systems are structured as time-limited courses or multi-session self-help programs: ACT concepts and exercises, including values clarification, are presented in fixed modules or sessions, and engagement and completion challenges are common, with repetition and length frequently cited as barriers.

GROW explores a complementary, non-clinical design space. Instead of offering a course-like intervention or treatment program, it uses ACT's values and committed action processes as design resources for everyday goal support. BEVS-inspired values reflection is integrated directly into the conversational interface and dashboard, rather than treated as a one-off assessment. Values are discussed through brief dialogue, used to frame goal setting, and revisited during check-ins when students encounter difficulty, while visual summaries on the dashboard support ongoing reflection. By embedding values-based reflection across lightweight, recurring interactions and linking it to concrete goals, reminders, and progress tracking within students' existing academic tools, GROW aims to support committed action over time without requiring extended sessions or a clinical framing.

## 2.3 SMART Goals Framework

Goal setting has long been shown to support motivation, self-regulation, and psychological well-being [36, 51]. The SMART goals framework, which encourages goals that are Specific, Measurable, Attainable, Realistic, and Timely, provides a widely adopted structure for translating intentions into actionable plans. Studies with university students indicate that SMART-based interventions can improve goal attainment, increase psychological need satisfaction, and support well-being [2].

Across many digital well-being systems, goals are commonly used to frame initial activities or personalize content, while subsequent interactions focus primarily on reflection, monitoring, or exercise completion rather than revisiting how goals evolve over time [24, 33, 68]. While effective for short-term support, this structure can limit opportunities for users to renegotiate goals as their priorities, constraints, or emotional states change.

From a theoretical perspective, Self-Determination Theory suggests that sustained well-being depends not only on goal achievement, but on whether goals continue to support autonomy, competence, and meaning over time [71]. This highlights a limitation of systems where structured goal frameworks such as SMART emphasize specificity and actionability but do not account for whether goals remain personally meaningful. Work in Acceptance and Commitment Therapy (ACT) emphasizes values-aligned action, but does not prescribe how such values should be translated into concrete, trackable goals. Together, these perspectives point to a design gap: existing systems rarely integrate values-based reflection with structured goal operationalization over time. GROW addresses this gap by treating SMART goal setting as a continuing conversational process rather than a one-time configuration step, while grounding goals in values elicited through dialogue. Goals are introduced and set through interaction, revisited during recurring check-ins, and can be revised, paused, or reframed in response to students' changing contexts. By coupling structured goal criteria with values-based reflection and longitudinal visualization, *GROW* supports goal pursuit as an adaptive, values-aligned process embedded within everyday routines rather than a static feature focused solely on completion.

## 3 The *GROW* System

*GROW* is a conversational AI system designed to support university students' well-being through structured goal setting, reflection, and values-aligned action. The system (Figure 1) combines an LLM-powered conversational coach, a reflective dashboard, and lightweight calendar and email integrations to embed practices that promote well-being (e.g. goal review, brief check-ins, and reflective prompts) into students' existing digital routines. Rather than replacing counseling or therapy, *GROW* is designed as a non-clinical support tool that complements university services by offering private, flexible, and ongoing coaching focused on self-awareness, motivation, and follow-through. It is not designed to diagnose, treat, or respond to crises; safety constraints and escalation pathways are described in Section 3.6.

### 3.1 Design Goals

Four design goals guided the development of *GROW* and inform the design of both the conversational coach and dashboard.

(1) **Treat goal setting as an evolving process.** Students' goals, priorities, and constraints change over time, particularly in academic contexts marked by fluctuating workloads and stressors [39, 54]. The system therefore needed to support revision, reflection, and recommitment rather than only one-time goal declaration and tracking. In *GROW*, students can define multiple goals, refine or downscope them in conversation, pause or extend them on the dashboard, and use check-ins to renegotiate commitments as circumstances shift. The design treats goals as living commitments that can be revisited rather than fixed contracts.

(2) **Connect reflection to concrete action.** Prior work notes that reflection systems vary in how they support action, and that a common barrier is the absence of specific, actionable guidance to help users apply reflective insights [6, 47]. *GROW* was designed to link reflective moments, such as values clarification or emotional check-ins, to concrete, time-bound goals and next steps. To achieve this, it draws on SMART goal setting [21] and Acceptance and Commitment Therapy (ACT) [41] as interaction design resources rather than therapeutic protocols, so that reflective dialogue flows into negotiated goals, plans, and follow-up prompts. This design treats reflection and action as a continuous, iterative process. In practice, this means that values discussed with the coach are explicitly used to anchor "why" a goal matters, and reflections on difficulty are followed by concrete adjustments to scope, timing, or strategy.

(3) **Integrate goal work into existing routines.** Prior work on digital mental health and well-being tools shows that engagement often declines when interactions require additional time, effort, or sustained attention outside users' everyday routines [58, 80]. *GROW* instead connects goal-related reflection and follow-through to tools students already rely on, such as calendars and email, by scheduling goal check-ins and sending reminders at times aligning with existing commitments (for example, before study blocks or after classes) rather than at random times. The aim is to make engaging with goals and reflection feel like a natural extension of students' current digital practices, rather than an additional app they must remember to open.

(4) **Respect non-clinical boundaries and student autonomy.** Prior work on behavioral intervention technologies, relational agents, and recent LLM well-being systems emphasizes the importance of clear role definition, expectation management, user autonomy, and appropriate boundary-setting in systems that engage with health, well-being, or behavior-change-related content [10, 40, 58]. This work cautions against treating user engagement as evidence of clinical benefit, and highlights the need for systems to make their scope and limitations explicit to avoid misplaced expectations or reliance.
*GROW* is intentionally designed as a non-clinical, values-based well-being coaching system rather than a therapeutic or diagnostic tool. The system does not assess symptoms or provide clinical advice. Instead, it focuses on supporting reflection, goal-setting, and follow-through while allowing students to decide when and how deeply to engage. Professional and institutional support resources are made persistently visible through the dashboard to reinforce role clarity and to encourage appropriate help-seeking beyond the system when needed.

We next describe how these design goals are instantiated in the conversational coach (Section 3.2), reflective dashboard (Section 3.3), and workflow integrations (Section 3.4).

### 3.2 Conversational AI Coach

The conversational coach serves as the primary interaction space in *GROW*. It is implemented as a prompt-engineered LLM configured to act as a goal-oriented coach rather than a therapist. The coach guides students through values reflection, goal definition, recurring check-ins, and adaptive coaching conversations. It is presented as an open chat that students can access at any time. The coach does not initiate free-form conversations autonomously; instead, it responds when students send messages, including when they return in response to calendar or email reminders. This behavior is intended to respect student autonomy and avoid unsolicited interruptions.

Interaction with the coach unfolds across three recurring phases: an introduction phase, a goal setting phase, and an active coaching phase. These phases are not strictly linear. Students may revisit values reflection while actively pursuing goals, or revise goals as their circumstances change, aligning with the design goal of treating goal setting as an evolving process.

***Introduction Phase.*** The introduction phase is designed to establish rapport and shape how the coach engages with the student over time. The prompt used in this phase is provided in Appendix A.1. Through an initial conversation (Example in Figure 3), the coach gathers basic contextual information (e.g., name, academic major, year of study) and invites reflection on broad personality tendencies using the Big Five (OCEAN) model, which captures patterns related to Openness, Conscientiousness, Extraversion, Agreeableness, and Neuroticism [76]. Students are also asked about their typical coping tendencies when facing stress or setbacks.

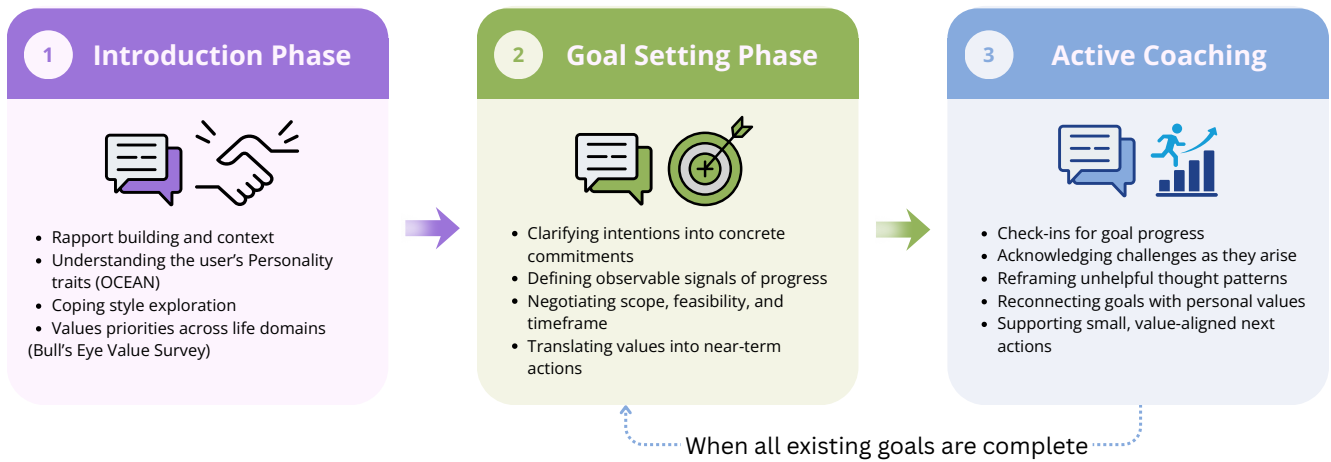

Figure 2: Interaction phases of the GROW conversational coach. The system supports students through an introduction phase (rapport, context, coping, and values reflection), a goal-setting phase (clarifying intentions into concrete commitments), and an active coaching phase (recurring check-ins, values-oriented reflection, and adaptive next actions).

These inputs are used to adapt subsequent interactions at a high level, including conversational tone, response length, and the balance between encouragement, reflection, and planning. The goal is not to assess mental health or label users, but to shape early interactions so they feel personally attuned and respectful of different communication and coping styles. For example, students who express higher openness may receive more exploratory prompts, while those with higher conscientiousness may be guided toward more structured, plan-oriented goal formulations. This design choice is informed by prior work indicating that aligning feedback with users' personality traits can support higher engagement and more effective digital coaching [28, 73]. Students can also request changes to the coach's style directly (for example, asking for shorter responses), and the coach is instructed to honor such preferences.

As part of this phase, students also complete a conversational version of the Bull's-Eye Values Survey (BEVS) [52]. The BEVS invites reflection on four life domains: work or education, relationships, personal growth or health, and leisure, by prompting students to identify personally meaningful values and consider how closely their recent actions align with them. The prompt used for this exercise is included in Appendix A.4. These value statements serve as an anchor for later goal-setting and coaching conversations, helping connect reflection to subsequent action.

*Goal Setting Phase.* Following the introduction phase, the conversational coach transitions into goal-setting conversations grounded in the SMART framework. Drawing on users' prior reflections and Bull's-Eye Values Survey (BEVS) responses, the system surfaces relevant value domains from earlier interactions to provide context, while leaving goal selection entirely user-driven. Rather than presenting a static form, the coach engages students in an interactive dialogue structured around the SMART goal framework. Through conversation (as shown in Figure 4), students are invited to clarify what they want to achieve (*Specific*), how progress will be recognized (*Measurable*), what feels feasible given current constraints (*Achievable*), why the goal matters to them (*Relevant*), and the intended duration or timeframe (*Time-bound*). This conversational approach allows goals to be refined in situ. As students articulate their intentions, the coach supports adjustment of scope, timelines, or expectations, enabling goals to evolve as clarity develops rather than requiring upfront precision. For example, when a student expresses a general intention such as "managing stress," the coach supports translating this intention into a concrete and time-bound commitment, such as practicing a short breathing exercise before study sessions each day for one week. The prompt used during this phase is provided in Appendix A.2.

Students may define multiple goals during this phase. Each goal is broken down into explicit, trackable components and is automatically surfaced on the dashboard to support ongoing review and reflection. By linking reflective insights (for example, values and stressors) to specific commitments and visible progress indicators, this phase directly operationalizes the design goal of connecting reflection to concrete action, while also supporting goal evolution over time.

*Active Coaching Phase.* Once goals have been established, the coach adopts an active coaching role focused on follow-through, reflection, and adaptation. This phase combines recurring check-ins with values-oriented reflection to help students sustain commitments while remaining responsive to changing circumstances.

Active coaching interactions are informed by principles from Acceptance and Commitment Therapy (ACT), which emphasize psychological flexibility and values-consistent action [30]. Rather than delivering structured therapeutic exercises, ACT's six core processes are used as interactional guidance for shaping brief, reflective prompts embedded within everyday dialogue. These prompts support acknowledgment of difficult experiences, stepping back

| | |
|---|---|
| **Coach:** Hi! I'm your wellbeing coach. How are you feeling today? I'm here to offer you emotional support and help you achieve your goals. I will be asking you a few questions to get to know you better - these are just a one time thing! ❶ | **Coach:** Let's do a quick values check-in. When you think about your studies, what matters most to you? ❸ |
| **Student:** Hi! I am feeling good. | **Student:** Being successful and receiving good grades. |
| **Coach:** Glad you're here. Can I know your first name, your college year, and what you're studying? | **Coach:** Thanks, Miya! Wanting to be successful is a strong value for Work/Studies. On a scale of 1 to 7, how close do you feel your current actions are to that value of "being successful" in your studies? (1 = not close at all, 7 = very close) |
| **Student:** I'm Miya, a senior studying Computer Science. | **Student:** I would say a 3. |
| **Coach:** Computer Science sounds fun. Now, just getting a sense of your personality: Are you the type who gets excited about trying new things and exploring ideas, or do you prefer sticking to what you know? ❷ | **Coach:** Thanks for being honest, Miya. Let's move to the next area: Relationships. When you think about friends, family, or people you care about, what kind of person do you want to be, or what matters most to you there? |
| **Student:** I love trying new things. | **Student:** I want to be able to put in effort and make time for my people. |

❶ **Introduction and Rapport Building:** Initial conversation to establish context and set expectations for the coaching interaction.
❷ **OCEAN Personality Traits and Coping Tendencies:** Asking briefly about personality traits in a friendly tone to understand broad personality tendencies and coping mechanisms that shape communication style.
❸ **Bull's Eye Values Survey:** Administering this conversationally to reflect on personally meaningful values across key life domains and current alignment.

Figure 3: Introduction phase excerpts in *GROW*. The figure illustrates how the conversational coach establishes context and personal grounding during the Introduction Phase. The interaction begins with brief rapport-building and collection of basic academic context, followed by lightweight prompts about broad personality tendencies using OCEAN framework and coping styles. The phase concludes with a conversational administration of the Bull's-Eye Values Survey, prompting reflection on personally meaningful values and perceived alignment across key life domains

| | |
|---|---|
| **Coach:** Now that we've looked at what matters to you, let's set one small goal you'd like to focus on. Which area feels most important right now: stress, sleep, study rhythm, or relationships? | **Coach:** Great. And how often would you like to do this in the next week? |
| **Student:** Stress relief. | **Student:** I think I want to do it everyday. |
| **Coach:** What's one small thing you could do this week that might help reduce your stress? | **Coach:** As a senior, there might be some busy or overwhelming days, what might help make this easier to follow through on those days? |
| **Student:** I could play my guitar. | **Student:** Placing the guitar near my bed. |
| **Coach:** Nice choice. To make it clearer, how long would you want to play each time? | **Coach:** Great that sounds practical. One last thing: why does this goal matter to you right now? |
| **Student:** About 10 minutes | **Student:** It really helps me calm down when I'm stressed. |
| | **Coach:** Amazing! We'll check in on this during the week, and you can always adjust it if needed. |

Figure 4: Goal-setting phase excerpts in *GROW*. Through a conversational flow grounded in the SMART framework, the coach helps students clarify what they want to do, how progress will be measured, what feels feasible, why the goal matters, and the intended timeframe. Chat excerpts are condensed to highlight how goals are shaped through dialogue rather than static forms.

from unhelpful thought patterns, and recommitment to personally meaningful actions, without framing anything in clinical or diagnostic terms.

During this phase, students initiate check-ins by reporting on goal progress, obstacles, or changes in circumstances, often in response to scheduled reminders in their calendar or email. The coach acknowledges progress, updates goal tracking, and tailors its responses to what the student describes. For example, as illustrated in Figure 5, when students report difficulty, low motivation, or self-criticism, the coach introduces brief ACT-informed reflections that normalize challenges, reconnect actions to personal values, or support adjustment of plans through smaller, value-aligned next steps. The prompt used during this phase is provided in Appendix A.3.

Reflections and updates are synchronized with the dashboard, allowing goals to be refined, paused, or extended over time. By allowing goal-setting and coaching to co-occur, the system emphasizes flexibility over completion pressure and supports sustained engagement by linking concrete actions to broader values and longer-term intentions. This coaching-oriented interaction style aligns with behavior change frameworks that emphasize reflective questioning and goal structuring to support agency and self-efficacy [57], as well as prior work showing that responsiveness and interpersonal attunement can strengthen engagement in conversational systems [10, 13].

## 3.3 Reflective Dashboard

The dashboard is a core component of the system that supports continuity across coaching interactions. It provides students with a view of their goals, engagement, and emerging patterns, helping them stay oriented and reflect on their progress over time rather than only within isolated conversations. It also links conversational activity to calendar-based reminders, making it easier to integrate goal work into existing study routines.

The dashboard is organized around a progress header and three tabs: **Support Resources**, **Goals**, and **Conversation Insights**. Each tab highlights different aspects of the student's ongoing interaction with the system, translating conversational activity into structured visual representations that support reflection, follow-through, and re-engagement.

*3.3.1 Progress Overview.* The progress header (Figure 6) shows a student's current phase in *GROW* (Introduction, Goal Setting, or Active Coaching), offering high-level orientation without requiring recall of prior conversations. Beneath this, two continuously updated indicators summarize engagement and follow-through:

- **Overall Goal Progress** reflects the average completion rate across all active SMART goals. It is computed by aggregating the completion percentage of each goal, and updating dynamically as users log progress or complete individual objectives.
- **Check-in Consistency** summarizes how frequently students have engaged with the coach in recent days. The indicator increases when students check in regularly and decays when there are longer gaps between interactions.

Both indicators are intentionally coarse and descriptive rather than precise or evaluative. Prior work in personal informatics suggests that lightweight progress cues can support reflection and sustained engagement without introducing pressure or performance framing [47]. In *GROW*, these indicators function as reflective signals that help users notice momentum and engagement patterns rather than measure success.

*3.3.2 Dashboard Tabs.* The dashboard includes three tabs:

*Support Resources Tab.* The Support Resources tab (Figure 7) provides a curated list of professional and institutional support resources. Links to campus counseling services, crisis helplines, and self-guided materials are presented with brief descriptions to support informed and easy access. Including professional resources directly within the dashboard reinforces role clarity by positioning *GROW* as a complementary, non-clinical support rather than a replacement for professional care. This design aligns with guidance from prior evaluations of digital mental health systems emphasizing the importance of clear pathways to external support [80].

*Goals Tab.* The Goals tab (Figure 8) displays the users' active and completed goals. Each active goal is displayed with its planned duration, next steps, and an estimate of current progress. Progress is inferred by the conversational coach based on user-defined, countable weekly measures and self-reported actions during check-ins. It is represented using a coarse indicator intended as a reflective cue rather than a precise or evaluative measure. Visualizing goal progress and consistency supports reflection on effort and engagement rather than precise performance; for example, students can quickly see which goals they are actively pursuing and which have stalled. From this tab, students can also see and adjust the optionally scheduled goal check-ins in their calendar, linking goal work directly to their existing routines.

*Conversation Insights Tab.* The Conversation Insights tab (Figure 9) shows higher-level patterns emerging from users' interactions with the coach. Rather than presenting full conversation histories, this tab provides concise summaries of recurring themes and descriptive communication patterns, such as tone or response style. These summaries surface **key themes, challenges, and moments of reflection from prior interactions**. They are designed to support introspection and to enable quick re-entry after periods of inactivity, helping students recall what they were working toward and why, without revisiting lengthy chat logs. Prior work in personal informatics shows that reflection is better supported when systems provide higher-level representations that help users make sense of experiences over time, rather than requiring review of detailed records [47]. Research on long-term conversational systems further highlights the importance of shared history and continuity cues in helping users resume interaction without reconstructing prior context [10]. In GROW, these summaries function as lightweight memory aids rather than complete or evaluative records of past conversations.

The dashboard also presents **high-level patterns in communication**, such as tone or message length across conversations. These patterns are not intended to characterize students or infer psychological states, but to provide a descriptive overview of how they tend to express themselves within the system. Prior work suggests that making behavioral patterns visible through aggregated representations can help users notice tendencies and prompt reflection [47]. In *GROW*, communication patterns are aimed to be reflective cues that support awareness of interaction habits over time.

Finally, the Conversation Insights tab includes a **dartboard-style values alignment visualization** derived from BEVS reflections as seen in Figure 9. The BEVS is a widely used ACT-based values exercise designed to make abstract priorities tangible and to surface discrepancies between values and actions without evaluative framing [8, 52]. In *GROW*, the dartboard visualization functions as a reflective prompt that invites consideration of how ongoing goals relate to personally meaningful domains. It highlights areas where actions feel closely aligned with values and areas where alignment is weaker, encouraging students to revisit or adjust goals accordingly, rather than serving as a measure of success or failure.

Together, these dashboard elements instantiate *GROW's* design goals by treating goals as evolving commitments, tying reflective

| | |
|---|---|
| **Student:** I'm here to check in. I played my guitar three times this week. | **Coach:** Noticing that thought is important. You don't need to push it away to keep going. Even with that pressure there, playing guitar still supports what you care about. Would it help to think of guitar time as a short reset rather than lost time? ❷ |
| **Coach:** Hi! Welcome! Thanks for checking in. I'm updating your progress now. You've completed 3 out of 7 planned sessions so far this week. How did it feel when you played? ❶ | **Student:** Oh yeah, well that feels better. |
| **Student:** It helped, but some days I was too stressed to start… | **Coach:** Got it. Would you like to keep this goal the same for the rest of the week, or adjust it to fit busy days better? ❸ |
| **Coach:** That makes sense. Stress can show up even when you're doing something helpful. When you felt that resistance, what was going through your mind? | **Student:** I want to keep it, even if I'm overwhelmed. |
| **Student:** That I should be working instead. | **Coach:** All set. I'll keep the goal as is and check in again soon. Even a few minutes still count toward staying consistent. You're doing great! |

❶ **Check-In and Goal Progress Update:** Initial conversation to establish context regarding goal progress for the coaching interaction.
❷ **Using Values and ACT's Core Processes:** When students describe difficulty or resistance, the coach invites reflection on internal experiences without evaluation, such as noticing pressure, stress, or self-critical thoughts, rather than attempting to remove them. Actions are reframed in relation to what the student cares about (e.g., stress relief, balance), helping reconnect the activity to its underlying purpose.
❸ **Goal Adaptation and Continuity:** Students are offered control to keep, modify, or adapt goals. Any updates are reflected immediately in the dashboard.

Figure 5: The student initiates a check-in by reporting recent goal-related activity ①, prompting the coach to acknowledge progress and update goal tracking. When the student describes difficulty or resistance, the coach responds with brief, ACT-informed reflections that invite noticing internal experiences without evaluation and reconnect the activity to personally meaningful values ②. The interaction concludes by asking the student if they want to keep or update the goal ③.

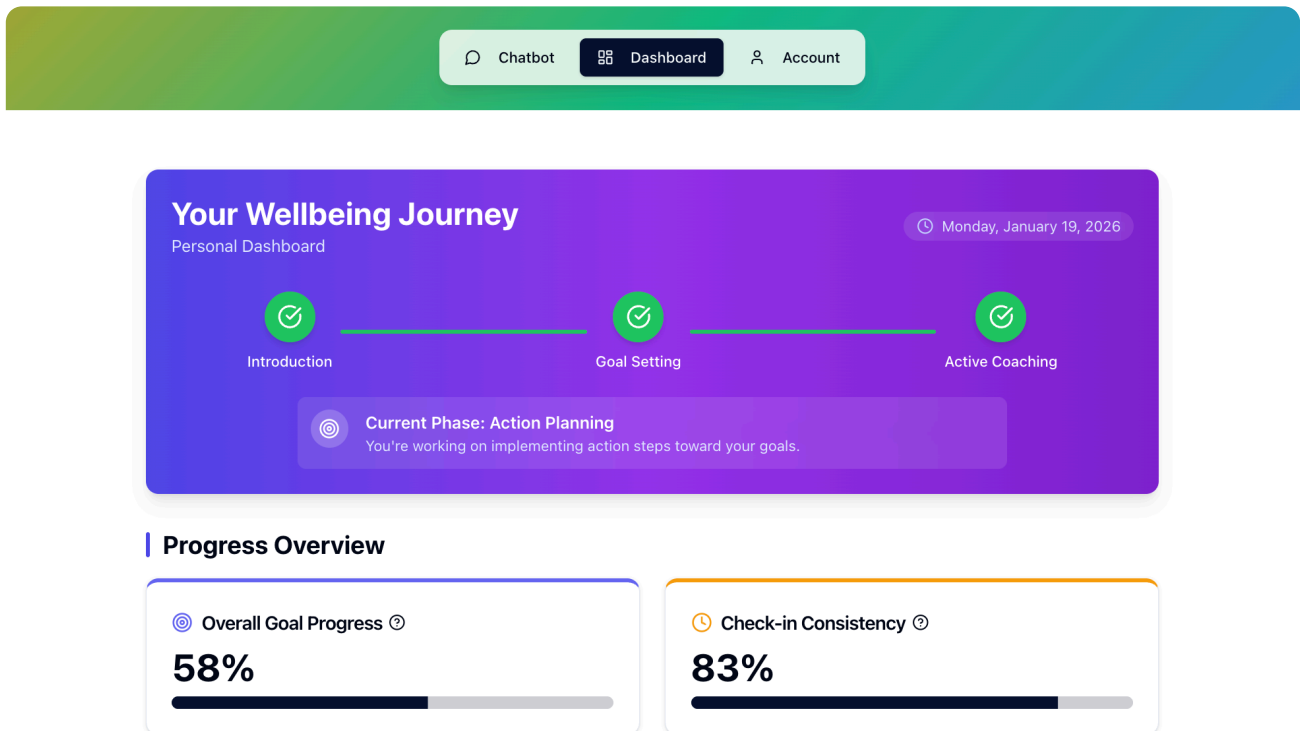

Figure 6: Progress header and overview section of the *GROW* dashboard. The header communicates the user's current stage in the coaching workflow (Introduction, Goal Setting, Active Coaching), while the overview presents lightweight indicators of overall goal progress and check-in consistency. Tooltips provide brief explanations of how each metric is calculated.

GROW: A Conversational AI Coach for Goals, Reflection, Optimism, and Well-Being insights to concrete actions and reminders, and embedding goal work within the digital routines students already use.

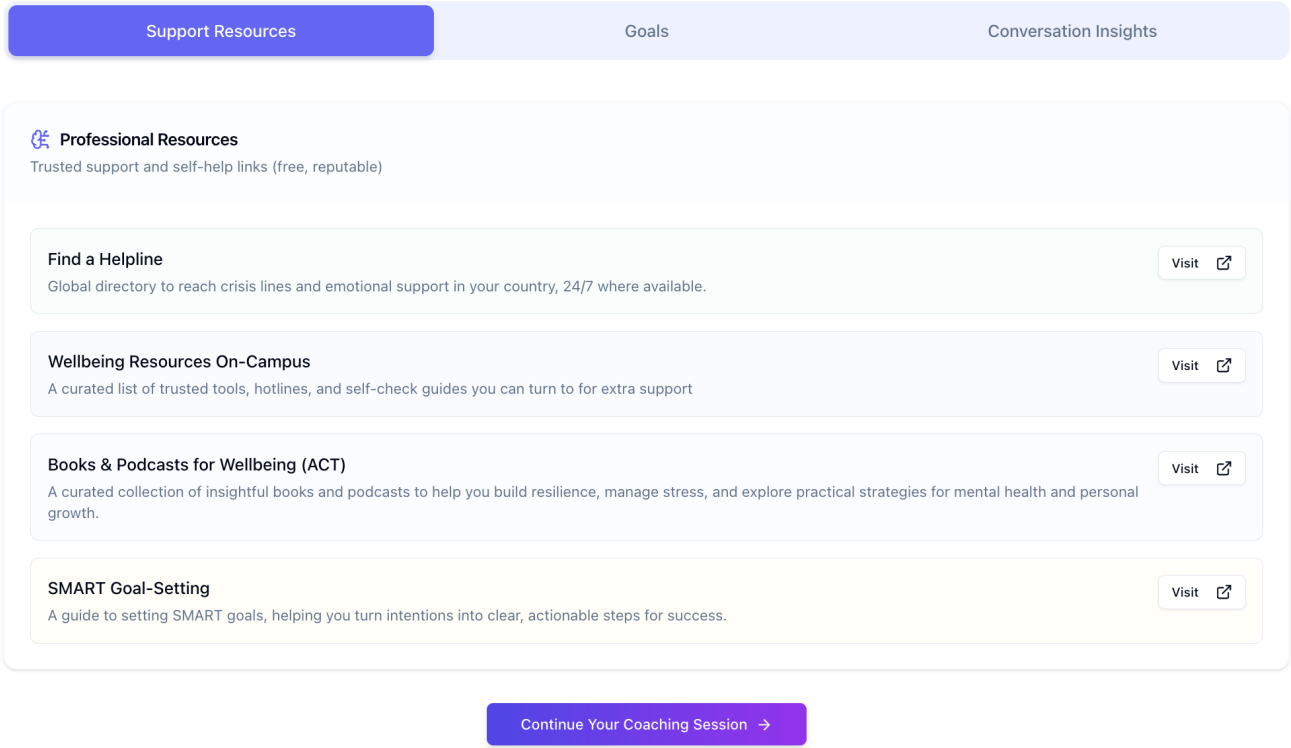

Figure 7: Support Resources tab in the *GROW* dashboard. The tab provides persistent access to external professional resources, including crisis lines, on-campus services, and evidence-based self-help materials, reinforcing that GROW is a non-clinical support tool rather than a replacement for professional care.

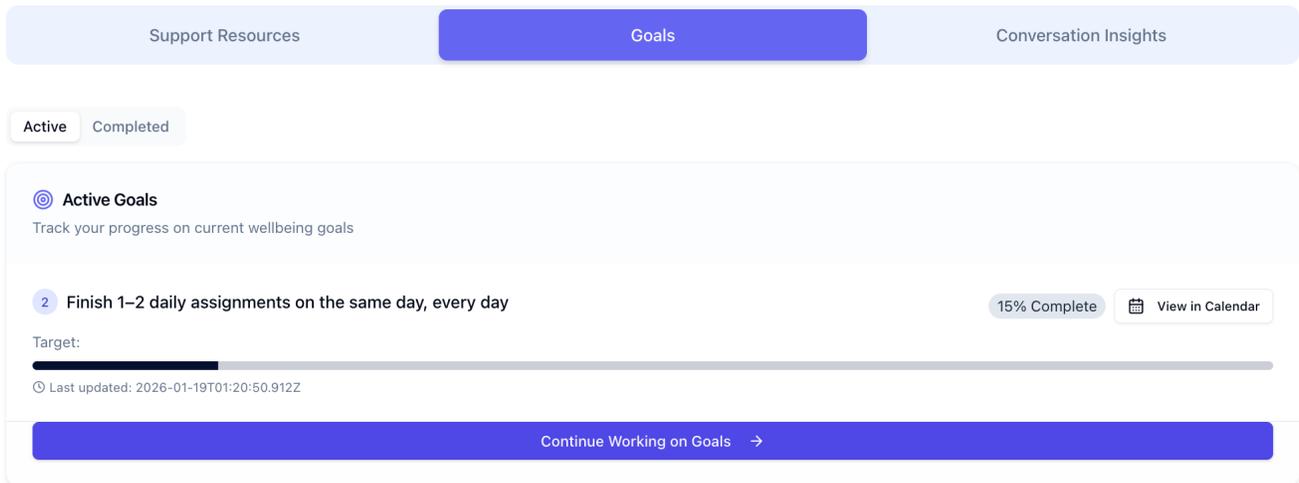

Figure 8: Goals tab in the *GROW* dashboard, showing students' active and completed goals along with associated duration, next steps, and coarse progress estimates derived from interaction history. Goals can be optionally added to students' personal calendars with scheduled check-ins and planned steps, or viewed in calendar from this tab once added.

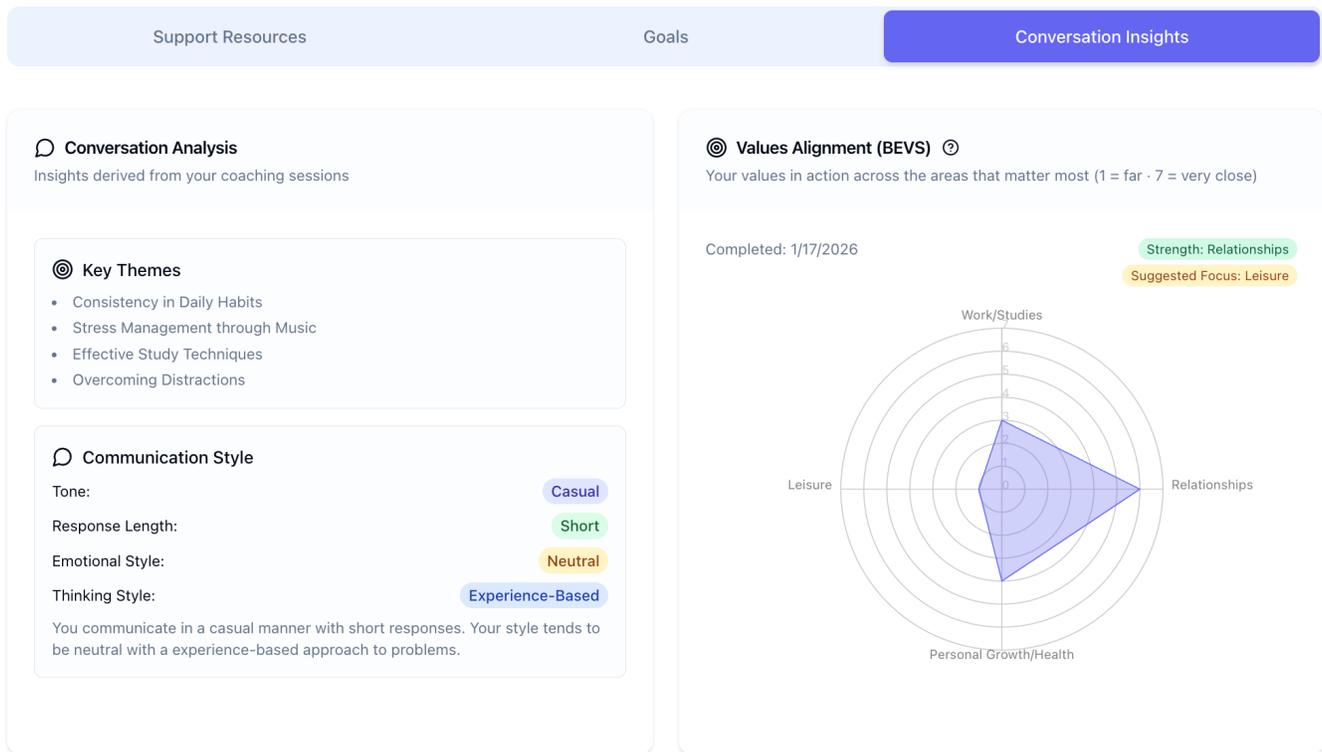

Figure 9: The Conversation insights tab presents high-level reflections derived from prior conversations, including recurring themes that come up in conversations with the coach, communication patterns (e.g., tone and response style), and a dartboard-style visualization of values alignment based on the BEVS taken conversationally.

### 3.4 Workflow Integrations

To support continuity beyond the chat interface and promote sustained motivation and follow-through, *GROW* integrates with tools that students already use in their daily academic workflows. Rather than positioning goal work as a separate activity, these integrations are designed to gently resurface goals and reflections at moments that align with students' existing routines.

*GROW* integrates with users' calendars to schedule lightweight goal-related check-ins when connected. When a goal is created, the system automatically adds check-in events in the students Google Calendar (as shown in Figure 10), typically at the midpoint and near the end of the goal's duration. Students can specify preferred time windows (morning, afternoon, evening, or night), and the system schedules events based on calendar availability to minimize disruption and avoid conflicts (Figure 11).

In addition to calendar events, students may opt into asynchronous email reminders that invite them to return to the system and reflect on active goals. Reminder frequency can be customized (daily, biweekly, or weekly, with weekly set as the default). These reminders are intended to support continuity during busy periods without enforcing rigid schedules or creating pressure to respond immediately.

Beyond scheduling preferences, students can personalize aspects of the conversational coach's identity. Users may optionally select a name, avatar, and preferred gender for the coach; if no changes are made, the coach remains in its default configuration. These options are intended to support comfort, familiarity, and approachability during interaction, allowing students to engage with a coach persona that feels personally acceptable or relatable.

### 3.5 Technical Implementation

*GROW* is implemented as a web-based conversational system that uses the OpenAI GPT API [61] as its language generation engine. The system maintains an explicit interaction state corresponding to the three coaching phases (Introduction, Goal Setting, and Active Coaching), which constrains both dialogue flow and data persistence across sessions.

User messages are processed synchronously in real time, and responses are generated with phase-specific prompting logic. Beyond the high-level personalization established during introduction phase, the system adapts response structure based on surface conversational cues such as expressed uncertainty, frustration, or reported progress. For example, inputs signaling difficulty or overwhelm trigger shorter, acknowledgment-oriented responses, while reports of progress or goal completion elicit more elaborative, planning-oriented feedback. To ensure consistency across interactions, the system employs few-shot prompting to establish a stable coaching voice and prompt chaining to manage transitions between interaction phases. Separate prompt templates (shown in A.1 - A.5) are used for Introduction, Goal Setting, Active Coaching, and the Bull's-Eye Values Survey, allowing the system to advance interaction state while keeping backend logic hidden to users. Model parameters (temperature and maximum token length) were adjusted to balance linguistic variability with predictability, for responses to remain supportive without drifting from coaching intent.

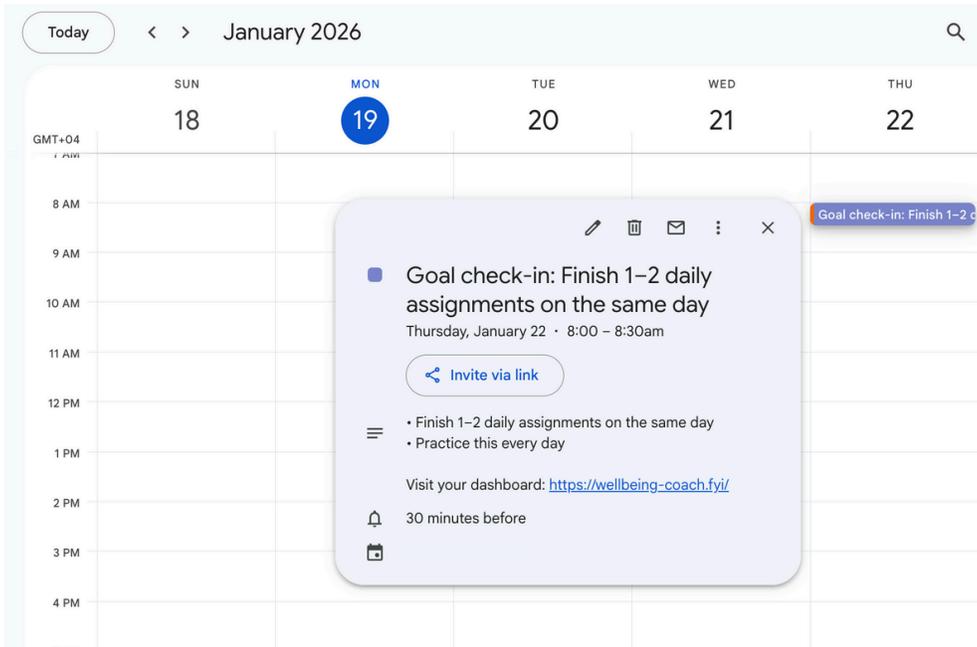

**Figure 10: A scheduled goal check-in appears in the student's calendar, linking directly to the *GROW* dashboard and reminding the student to reflect on progress toward an active goal. It also shows the decided steps. Two events for the goal will be scheduled at the user's preferred time slot: one at midpoint of planned duration and one at end.**

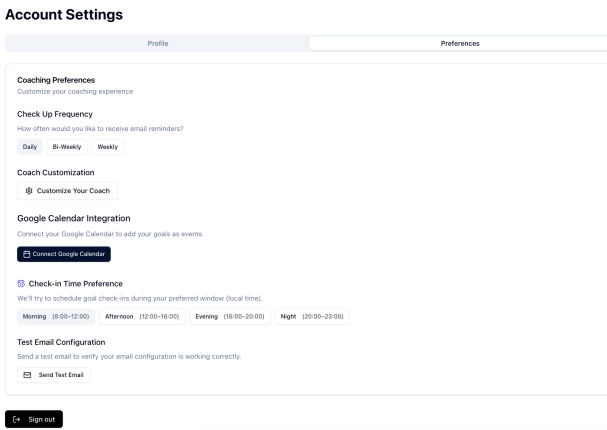

**Figure 11: Settings screen allowing users to customize coaching preferences, including the frequency of email reminders and preferred time windows for calendar-based check-ins.**

Structured user data are persisted through a dedicated tool, `saveProfile`, rather than storing full conversation transcripts. Model outputs intended for persistence are validated against a Zod-based schema (`ProfileJsonSchema`) and merged into a MongoDB [59] user record via Prisma [64]. Each tool call is tagged with the active interaction phase, restricting writes to phase-relevant fields and preventing unintended overwrites. Additional safeguards include schema validation, duplicate-write detection, and checks that prevent re-saving completed profile sections.

### 3.6 Ethical and Privacy considerations

*GROW* was designed with a privacy-preserving and ethically responsible framework to ensure that student well-being is supported without compromising autonomy, consent, or confidentiality. Several aspects of the system's design and deployment explicitly reflect these considerations.

**Privacy-Preserving Data Architecture.** User inputs are sent to the LLM to generate responses, but system-level PII (e.g., email and account details) is handled exclusively by Clerk, which manages authentication and stores user identity within its own secure, encrypted infrastructure. This information is not included in model requests or the application database. During interaction, user-provided names may be used transiently in prompts (client-side) to support natural dialogue, but are not stored server-side.

All stored data is linked only to a pseudonymous user identifier (Clerk user ID). Conversational logs are retained to support goal tracking and continuity across sessions, with PII removed prior to storage where possible. No raw identifiers are accessible to the system's reasoning components. Third-party integrations follow a least-privilege model: for example, calendar access is restricted to availability (free/busy) metadata, and event content is neither read nor stored. This separation ensures that sensitive personal data remains isolated from interaction-level processing.

**Boundaries of Support, Role Clarity and Safety.** Although *GROW* integrates principles from Acceptance and Commitment Therapy (ACT), it is not positioned as a clinical or diagnostic tool. It offers structured coaching and values-based reflection, not therapy, and professional and institutional support resources are made persistently available through the dashboard's Support Resources tab. The Support Resources tab includes prominently placed links to on-campus counseling, crisis lines, and evidence-based mental health resources. The system does not perform automated detection or classification of user distress. In cases where users explicitly reference distress in their input (e.g., self-harm related language), the system does not endorse or validate harmful actions and instead responds by encouraging users to seek professional support and redirects them to these resources that are permanently available on the dashboard. This design choice is intended to reinforce role clarity and reduce the risk that users perceive the system as a replacement for licensed care. However, as a conversational system, *GROW* may still be perceived in relational terms by some users (e.g., as a "friend"), even without explicit design intent. While the system is framed as a structured coaching tool, this highlights an inherent risk of personification in such systems.

**Transparency.** Throughout the dashboard, tooltips and labels explain how features such as goal progress, summaries, and visual reflections are generated from conversational input.

**Future Ethical Safeguards.** As *GROW* evolves, we plan to incorporate ongoing ethical auditing of language generation (e.g., checking for overly directive, biased, or emotionally misaligned responses) and user-controlled data deletion features. We will focus on mitigating unintended personification by introducing constraints on relational language, clearer system framing, and interaction designs that reduce the risk of over-reliance or perceived social substitution. These improvements aim to further align the system with emerging standards in AI ethics, user agency, and responsible well-being technology.

## 4 Formative Practitioner Study

This phase aimed to gather expert feedback on the system's features, conversational flow, and dashboard design to ensure theoretical and practical soundness before student testing.

### 4.1 Participants

5 licensed professionals, including 3 clinical psychologists (1 of whom is also a professor of psychology), 1 doctor, and 1 student-success coach, participated. All had prior experience in student mental health or digital well-being interventions. Sessions were conducted remotely via Zoom and lasted approximately 60 minutes.

### 4.2 Protocol

After providing informed consent, participants were asked about their professional backgrounds and prior experience with student well-being or digital intervention tools. The researcher introduced the study aims and clarified that *GROW* is a non-therapeutic coaching system focused on goal setting and well-being rather than a replacement for therapy.

Participants were then provided with a link to access the app. The walkthrough portion of the session took approximately 30 minutes, with the remaining time dedicated to discussion and interview questions. Using the concurrent think-aloud protocol [22], they were asked to share their screens while completing a series of tasks and to verbalize their thoughts during navigation. This approach allowed the researchers to observe real-time interactions and collect spontaneous feedback, while providing only minimal guidance to preserve natural discovery and professional judgment. The session consisted of the following tasks:

- **Task 1: Sign-In.** Participants signed in to the system and explored the landing interface.
- **Task 2: Introductory Interaction.** Participants interacted with the chatbot as it collected background information, including the Big-Five Personality Questions and BEVS, to simulate personalized coaching and values-based reflection.
- **Task 3: Goal Setting.** Participants were asked to create a personal goal using the chatbot, which guided them through refining it into a SMART goal.
- **Task 4: Active Coaching.** Participants performed a short check-in with the chatbot to experience how the system incorporates ACT principles within conversation flow and delivers feedback, encouragement, and reflection prompts based on progress.
- **Task 5: Calendar and Email Integration.** Participants connected Google Calendar, explored automated check-in scheduling, and adjusted email reminder frequency in the Settings tab.
- **Task 6: Dashboard Review.** Participants reviewed all aspects of the dashboard. The dashboard contained data and visualizations that were generated after participants' interactions with the chatbot (from Task 2-4)

After each task, participants provided open-ended reflections on what worked well, what could be improved, and how the features might be used in a university setting. The session concluded with a semi-structured interview guided by open-ended questions on overall usability, satisfaction, adoption intent, and perceived appropriateness, with follow-ups based on issues raised during the walkthrough. Participants were also asked to complete a short quantitative survey of 5-point Likert-scale questions assessing individual system features with response options ranging from *Not at all useful* to *Extremely Useful* and the System Usability Scale (SUS) [4]. All sessions were audio-recorded and transcribed for later analysis.

### 4.3 Data Analysis

Data from the formative study was analyzed using an inductive semantic thematic analysis [12]. Two authors independently reviewed the transcripts and generated initial codes by labeling explicit statements, questions, and reactions expressed by participants during the semi-structured interview. Coding focused on experts' articulated

GROW: A Conversational AI Coach for Goals, Reflection, Optimism, and Well-Being perceptions of system clarity, interaction flow, ethical boundaries, and suitability as a non-clinical well-being support, rather than on predefined theoretical constructs. Codes were then compared and iteratively grouped into higher-level themes that captured recurring patterns across participants. Themes were refined through discussion between the two authors, with disagreements resolved through consensus. The resulting themes were used to identify design-relevant insights that informed revisions to the system prior to the student deployment.

### 4.4 Findings

Experts rated G*ROW* as highly usable overall, with a mean System Usability Scale score of 83.5 ($SD = 5.2$), placing the system in the "Excellent" usability range according to established benchmarks [4]. The findings below summarize experts' perceptions of *GROW's* design, highlighting strengths, concerns, and opportunities identified prior to student deployment. We report themes that reflect experts' perspectives on the system's role, interaction design, and appropriateness for a university well-being context, supported by example quotes.

***Positioning GROW as Complementary Support.*** Experts consistently emphasized the importance of positioning *GROW* as a complementary, non-clinical support rather than a replacement for professional care. They described the system as an accessible, low-pressure space for reflection and accountability, particularly valuable for students who may be hesitant to seek counseling. One expert highlighted the importance of sequencing support appropriately, noting that "having that professional first step of things is really important to triage first, and then having, obviously this as an accountability partner is a really, really, I think, a strong tool" (P1). Others emphasized the system's availability and reach, especially outside formal service hours. As one participant noted, "anything that makes useful support services available to students, especially something, you know, this is available 24/7" (P4). Another expert reflected on its potential impact for a specific student population, stating, "I definitely think it would be beneficial for college students. I do think that there is a population of students that this will be really helpful for. I would recommend that students utilize it" (P2).

At the same time, participants cautioned against over-reliance and highlighted the need for clear boundaries, including introduction phase that communicates the system's scope and directs users toward professional resources when appropriate. This positioning was reflected in experts' positive evaluations of features that emphasized guidance and reflection rather than intervention, including values-based reflections and progress tracking (Table 1, Table 2).

***Grounding Reflection in Evidence-Based Frameworks.*** Participants viewed the system's grounding in ACT and SMART goal setting as a key source of credibility and structure. They confirmed that these frameworks helped legitimize *GROW* as more than a generic chatbot. One expert described the values-first approach as particularly compelling, saying, "I really like the value approach, especially starting with that kind of exploring areas of value" (P1).

Several participants highlighted how small, structured steps helped make progress feel tangible. One expert explained that seeing progress reinforced motivation, observing that "even though I am doing things that are small, it is making a difference... it provides some level of like reinforcing my good behavior or my tasks or the goals that I'm working on achieving" (P2). Another described how the system already supports breaking down overwhelming goals, noting that it suggests "just a list of steps, which are very small, which don't seem overwhelming or boring" (P3).

Table 1: Phase 1: Practitioner evaluations for chatbot elements on a 5-point Likert scale (N=5).

| Feature | Mean ± SD | Median | Mode |
| --- | --- | --- | --- |
| Personalized communication style | 3.8 ± 1.09 | 4 | 4 |
| SMART goal setting | 4.6 ± 0.55 | 5 | 5 |
| ACT principles | 4.2 ± 0.84 | 4 | 4 |
| Bull's Eye Value Survey | 4.2 ± 0.84 | 4 | 4 |

At the same time, experts suggested that brief, accessible explanations of ACT concepts would help students understand the intent behind reflective prompts. As one participant noted, "I think it's a good model. But so far we haven't had any explanation as to what ACT is... whether you add a little question mark bubble here, giving an overview of ACT" (P4). Consistent with these perspectives, SMART goal setting ($M = 4.50$) and ACT-informed components ($M = 4.20$) received among the highest usefulness ratings among chatbot features (Table 1).

***Designing for Engagement Without Clinical Claims.*** Experts consistently highlighted the importance of a warm, conversational tone that fosters accountability without feeling clinical or evaluative. They emphasized that engagement depended less on novelty and more on clarity, relatability, and sustained motivation. Visual simplicity and lack of cognitive overload were repeatedly mentioned as strengths. One expert noted, "it's very clean, it's very neat... people aren't going to feel overwhelmed by this" (P4). Suggestions included clarifying that introduction phase questions are a one-time activity, improving visual readability, and supporting flexible reminders. These recommendations reflected a broader concern with ensuring the system feels supportive and easy to return to over time, rather than demanding or burdensome.

Feature-level ratings supported these observations, with communication style summaries ($M = 4.67$), conversation insights ($M = 4.50$), and goal progress tracking ($M = 4.33$) rated highly among dashboard elements (Table 2).

***Safety, Privacy, and Ethical Use as Preconditions for Adoption.*** All participants stressed that trust was extremely important in student well-being contexts. They emphasized transparency around data use, optional demographic questions, and clear communication of privacy practices. Experts viewed the system's anonymized data handling as essential for user safety and recommended making privacy guarantees explicit during onboarding. Some participants also discussed the potential value of aggregated insights at an institutional level, provided individual anonymity is preserved.

### 4.5 Design Iterations

Feedback from the formative expert study informed a set of design changes made prior to student deployment. These refinements

Table 2: Phase 1: Practitioner evaluations for dashboard elements on a 5-point Likert scale (N=5).

| Feature | Mean ± SD | Median | Mode |
|---|---|---|---|
| Goal progress tracking | 4.4 ± 0.55 | 4 | 4 |
| Emotional well-being trend | 4.4 ± 0.55 | 4 | 4 |
| Check-in consistency tracker | 4.0 ± 1.00 | 4 | 3 |
| Links to professional resources | 4.4 ± 0.89 | 5 | 5 |
| Key themes | 4.6 ± 0.55 | 5 | 5 |
| Communication style summary | 4.8 ± 0.45 | 5 | 5 |
| Bull's Eye values visualization | 4.6 ± 0.89 | 5 | 5 |

clarified the system's role and improved interaction clarity. Key changes are summarized below.

- **Clarifying system role and boundaries.** Experts consistently emphasized that *GROW* should be positioned as a complementary, non-clinical support tool rather than a replacement for therapy or counseling. In response, we made this explicit in the study onboarding communication email, clarifying the system's purpose, limits, and non-clinical nature. We also ensured that links to institutional and crisis support resources were clearly visible within the dashboard. These changes aimed to support trust, appropriate expectations, and responsible use.
- **Making psychological frameworks more legible in interaction.** While experts valued the system's grounding in ACT and SMART goal setting, they noted that explicitly naming psychological frameworks, jargon or techniques during interaction (e.g., referring to "ACT tips," "values alignment," or "defusion exercises") could feel unclear or overly technical for some students. In response, we avoided dropping framework labels in isolation and instead introduced brief, plain-language explanations when such concepts appeared, grounding them in the immediate conversational context.
- **Refining conversational flow and tone.** Experts highlighted the importance of maintaining a warm, non-judgmental conversational style that supports accountability without feeling clinical or prescriptive. Based on this feedback, conversational prompts were revised to reduce perceived pressure, clarify that onboarding questions are a one-time interaction, and emphasize encouragement and reflection over evaluation.
- **Improving visual clarity and navigability.** Feedback also pointed to opportunities to strengthen readability and orientation within the dashboard. In response, visual contrast between tabs was increased, layout spacing was adjusted, and progress indicators were simplified to reduce cognitive load and support quicker sensemaking during brief check-ins.
- **Expanding and contextualizing support resources.** Experts recommended broadening the range of supporting materials available to students. As a result, the dashboard's professional resources section was expanded to include curated ACT-based books, podcasts, and self-guided materials alongside campus-specific services and crisis resources, enabling students to explore well-being practices beyond the chatbot interaction.

## 5 In-situ Student Deployment Study

This phase aimed to evaluate *GROW*'s real-world engagement, usability, and perceived impact on student well-being through a one-week deployment with undergraduate students.

### 5.1 Participants

35 undergraduate students (ages 18 and above) were initially recruited via campus mailing lists. All participants were screened using the Patient Health Questionnaire (PHQ-9) [37]; only those scoring below 15 (indicating mild to moderate symptoms) were invited to participate to ensure ethical safety and to avoid exposing clinically vulnerable individuals to a non-therapeutic tool. 4 participants discontinued before completing the second survey and 1 participant discontinued before the final survey. Hence, the final participant pool consisted of 30 students (10 female, 20 male): 6 freshmen, 4 sophomores, 10 juniors, and 10 seniors. All participants provided electronic informed consent through the IRB-approved portal before beginning the study. We did not explicitly collect measures of participants' prior experience with digital well-being applications or self-reported technology proficiency. As participants were recruited from a university population and interacted with a web-based system during the study, we assume a basic level of familiarity with digital tools.

### 5.2 Protocol

The study spanned seven days, with Days 1–4 designated as mandatory participation and Days 5–7 optional. After consent, participants received a link to the *GROW* application along with a demo video(approximately 2 minutes) that introduced the basic structure and navigation of the system. The video demonstrated the sign-in process, the layout of the main tabs (chat, dashboard, and settings), and how to connect Google Calendar and configure email reminders. Its purpose was limited to familiarizing participants with the interface and available features. The onboarding materials clarified that *GROW* functions as a well-being and goal-setting coach rather than a clinical or crisis-response system.

The week-long study followed the sequence outlined below:

- **Baseline Assessment (Day 1).** Participants first completed the PERMA-Profiler [15] to establish baseline well-being across the five domains of Positive Emotion, Engagement, Relationships, Meaning, and Accomplishment, as well as additional items on health, negative emotion, and loneliness. The PERMA-Profiler was selected for its brevity, psychometric reliability, and suitability for capturing broad well-being outcomes rather than disorder-specific measures. Participants then interacted with the chatbot's introductory phase and familiarized themselves with the dashboard.
- **Daily Use (Days 1 - 4).** Participants engaged in daily check-ins with the chatbot, setting or updating SMART goals, viewing progress visualizations on the dashboard, and receiving optional email reminders and Google Calendar notifications if they had integrated their accounts.
- **Mid-Study Evaluation (Day 4).** To assess interim well-being and usability, participants completed a second administration of the PERMA-Profiler and the System Usability Scale (SUS) [4].

- **Optional Use (Days 5 - 7).** App use during this period was made optional to observe whether participants would voluntarily continue engaging with *GROW*. Reminders were sent via automated email and calendar notifications based on the frequency and time set by the user.
- **Final Reflection (Day 8).** Participants completed a final mandatory survey evaluating how the chatbot and dashboard supported their goal progress and well-being. The survey included the PERMA-Profiler and a combination of quantitative and open-ended questions assessing specific system features, comfort, engagement, perceived impact, and a 5-point Likert-scale questionnaire assessing ease of navigation, visual appeal, and overall satisfaction, with response options ranging from Strongly disagree to Strongly agree.

### 5.3 Data Analysis

To understand how students experienced *GROW*, we analyzed open-ended survey responses and summarized results from standardized questionnaires (PERMA Profiler and System Usability Scale).

The open-ended responses from the final reflection survey were analyzed using an inductive, semantic thematic analysis approach. Analysis focused on students' explicit descriptions of how the chatbot and dashboard supported (or hindered) reflection, goal pursuit, motivation, and engagement over time. Two authors independently familiarized themselves with the survey responses and generated initial inductive codes by labeling recurring statements, concerns, and reflections expressed by participants. Codes were then compared and iteratively grouped into higher-level themes that captured shared patterns across participants' experiences. Themes refined through discussion and disagreements in interpretation were resolved through discussion until consensus was reached.

Quantitative data from the PERMA-Profiler and the System Usability Scale (SUS) were analyzed descriptively. Mean scores and standard deviations were computed to examine changes in self-reported well-being across the study duration and to assess perceived usability of the system.

### 5.4 Findings

Overall, participants reported that *GROW* was usable, supportive, and helpful for goal-related reflection and well-being. The system received a mean System Usability Scale (SUS) score of 77.5 ($SD$ = 5.2), placing it within the *Good–Excellent* usability range. Likert-scale responses further reinforced this assessment (Figure 12). Across all evaluated dimensions, the majority of participants selected *Agree* or *Strongly agree*. Ease of navigation received the strongest endorsement, with 97% of participants rating the system positively. Visual appeal and interface responsiveness were also rated favorably, with over 80% positive responses in each case. Overall design satisfaction similarly showed high agreement, indicating that students found the system both intuitive and pleasant to use.

#### 5.4.1 Conversational Coach.

***Perceptions of Responsiveness and Follow-Up.*** Students described *GROW* as most engaging when it appeared attentive to their goals and followed up on prior context. Remembering earlier conversations and referencing specific details helped create a sense

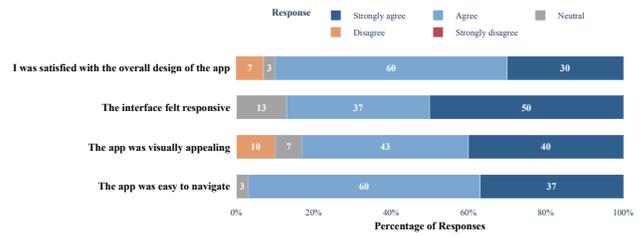

**Figure 12: Overall usability evaluation of *GROW*. Participants rated ease of navigation, visual appeal, interface responsiveness, and overall satisfaction on a 5-point Likert scale ranging from *Strongly disagree* to *Strongly agree*.**

of being "understood" and "remembered," which contributed to trust and connection. As one student noted that "It remembered that I had midterms and asked how I'm coping." (P4) while another described it "like talking to a friend who keeps me in check." (P16)

However, experiences of responsiveness varied. Some participants felt that the chatbot adapted to their context, while others described interactions as repetitive or static. For example, one student said that the system "set goals but did not adapt." (P17) and another felt that "It operated on the same level throughout the week." (P27)

Tone played an important role in shaping these perceptions. When responses felt warm and non-judgmental, students described the interaction as uplifting or genuine. One participant shared that "talking to it uplifts my mood" (P15). In contrast, repetitive phrasing reduced engagement, with students describing responses as "robotic" or "shallow" when variation and specificity were lacking (P23, P11).

***Comfort and Hesitations in disclosing to the Coach.*** Students reflected on how the non-human nature of the chatbot shaped comfort with self-disclosure. Many described feeling able to share personal experiences due to the perceived neutrality and lack of judgment like a student said "I'm fine sharing very personal things… because it is AI." (P9)

The absence of social presence reduced pressure associated with being evaluated or observed. However, despite transparency around data handling, some participants remained hesitant to disclose certain information. One student noted, "I knew it would keep it confidential, but I still didn't want to share everything." (P7)

In summary, while the chatbot's non-judgmental nature lowered barriers to disclosure for many students, privacy concerns continued to shape the depth of sharing for some.

***Perceived Helpfulness for Goals and Well-Being.*** Students rated the chatbot's perceived helpfulness positively across both dimensions. For goals support, 87% of respondents rated the chatbot as *Somewhat helpful* or *Very helpful* ($M$ = 4.17), with 30% giving the highest rating and 13% selecting a neutral option. For well-being support, 80% of respondents rated the chatbot as *Somewhat helpful* or *Very helpful*, 17% selected a neutral option, and only 3% rated it *Somewhat unhelpful*; no participants chose *Very unhelpful*.

#### 5.4.2 Dashboard: Supported Structure and Reflection.

***Structure, Progress Visibility, and Motivation***. Students described the system's structure and progress visualization as central to sustaining motivation. Seeing goal progress displayed on the dashboard helped make effort feel tangible and reinforced follow-through. One participant said that "check-in consistency made me more productive" (P29), while another noted that the dashboard "encouraged me to make an effort to achieve my goals" (P23).

Recurring check-ins helped break larger goals into manageable steps, transforming intentions into concrete actions. Calendar and email integrations further supported accountability for participants who enabled them, with students describing the system as something that "kept me on track" (P5).

Several participants suggested that lightweight notifications or spontaneous nudges could further support re-engagement during busy periods. In summary, structured check-ins combined with visible progress cues helped sustain motivation by making goal pursuit feel steady, manageable, and achievable.

***Self-Awareness through Reflective Feedback***. Participants highlighted reflective dashboard features, particularly *Communication Style* and *Values Alignment*, as meaningful tools for self-understanding. These visualizations prompted reflection on habits, priorities, and how goals aligned with what mattered to them. For instance, one student shared that the dashboard "made me think about how I am at a communicative level" (P13), while another noted it "helped me adjust my priorities" (P19).

At the same time, some participants wished the chatbot would offer more probing or challenging feedback rather than affirmation alone as "It always agrees with me… I wish it asked more questions." (P9)

This highlighted a tension between maintaining a supportive tone and offering feedback that feels sufficiently substantive to prompt deeper reflection.

***Perceived Usefulness of Dashboard Features***. Students' evaluations of the dashboard elements can be seen in Figure 13. Overall, participants rated all dashboard features positively, with Goal Progress Tracking ($M = 3.80$) and Check-in Consistency Tracker ($M = 3.63$) as the most useful components supporting reflection and progress. Key Themes ($M = 3.63$) and Communication Style Summary ($M = 3.60$) were also rated favorably, suggesting that students valued features that provided personalized insights from their interactions. Values Alignment Dartboard ($M = 3.47$) and Professional Resources ($M = 3.33$) received moderately positive ratings. These results indicate a generally positive reception of the dashboard, with higher ratings for features that directly supported goal reflection and progress tracking.

*5.4.3* **Patterns of Engagement Over Time**. Engagement patterns varied once participation became optional. Eleven participants continued interacting daily, five used the system intermittently, six interacted once or twice, and eight discontinued use after the mandatory phase.

Some students continued interacting daily or intermittently, often describing use as habitual or familiar. As one participant shared, "I think it became a habit of checking in with the chatbot before I would sleep" (P6).

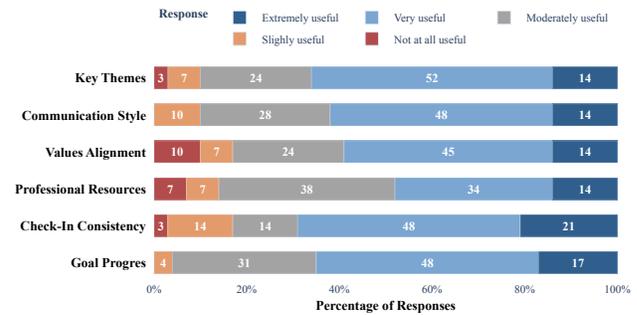

**Figure 13: Phase 2: Student Evaluation of the Dashboard Features.**

In contrast, reduced engagement was most commonly attributed to academic workload. One student said that they "Got super busy with midterms" (P17), while another noted that although the system was helpful, "I was very busy so I couldn't find time to talk to it everyday" (P29).

Others reported stopping once they felt they had achieved their goals, or due to the absence of stronger reminders prompting re-entry.

In summary, engagement followed two trajectories: sustained use among students who integrated brief check-ins into routine, and drop-off when competing academic demands increased or reminders were insufficient.

*5.4.4* **Changes in PERMA Profiler Outcomes**. Analysis of PERMA-Profiler scores showed modest changes over time. From Phase 1 to Phase 2, scores remained relatively stable ($M \approx -0.15$, $SD \approx 0.7$). Between Phase 2 and Phase 3, scores increased by approximately +0.45, with roughly 70% of participants showing net gains across the study.

Positive Emotion, Engagement, and Accomplishment showed the largest increases, while Health remained stable and Negative Emotion decreased slightly.

In summary, quantitative results align with qualitative accounts, indicating strong usability and perceived value, particularly for features supporting structure, reflection, and follow-through.

## 6 Discussion
## 6.1 Cross-Phase Synthesis of Expectations and Experiences

To relate formative expectations to real-world use, we synthesized expert and student perspectives across shared themes (Table 4). The table highlights where expert expectations aligned with, and where they differed from, what students reported. In general, experts and students pointed in similar directions, especially around the value of attentive follow-up, structured goal support, and flexible use. The synthesis also points to nuances in how reflective feedback was experienced and how engagement changed during busy periods, which we return to in our discussion of trust and continuity, the role of representations, and the use of well-being frameworks as background structure.

Table 3: Representative participant quotes.

| Findings | Participant quotes |
|---|---|
| Perceptions of responsiveness and follow-up | • "The memory it holds. I loved it when it remembers everything I said to it." (P23)<br>• "It followed up on what I said in our first conversations." (P22)<br>• "It kept me in check with my workload" (P4) |
| Perceived tone and emotional quality | • "I liked its positive tone. It kept encouraging me always, even when I said I forgot to do some things." (P5)<br>• "Somehow felt like a real interaction." (P22)<br>• "Some of the responses were getting repeated." (P15)<br>• "Sounded robotic so was hard to take it seriously." (P23)<br>• "Interactions felt really shallow." (P11) |
| Perceived lack of adaptation | • "Some of the responses were getting repeated in different instances." (P17)<br>• "It operated on the same level throughout the week." (P27) |
| Comfort and hesitations in disclosing to the coach | • "I liked that it gave me space to choose whether I wanted a deeper conversation or a short one." (P9)<br>• "Felt very secured using the app." (P20)<br>• "It's not judgmental." (P22)<br>• "I felt it was not judgmental. Very neutral takes" (P24)<br>• "It's a robot, it's not gonna judge or care" (P28) |
| Structure, progress visibility, and motivation | • "I found goal progress tracking so fascinating. It encouraged me to make an effort to achieve my goals." (P25)<br>• "Tt served as an active reminder of pending issues." (P4)<br>• "It helped me as a reminder." (P1)<br>• "Values alignment chart was particularly surprising and helped me adjust my priorities accordingly." (P21) |
| Patterns of engagement over time | • "I am still using it!" (P11)<br>• "I liked the chatbox feature so I continued telling my story." (P19)<br>• "I found the grounding exercises given by the chatbot to be very useful, particularly since it was a stressful exam week." (P21)<br>• "It's midterm season and I was swamped" (P16)<br>• "I continued because over time I thought it's like a normal online person rather than a bot. " (P30) |

Table 4: Cross-phase synthesis of expert expectations (Phase 1) and student experiences (Phase 2).

| Theme | Expert expectations | Student experiences |
|---|---|---|
| Responsiveness and trust | Emphasized importance of warmth, follow-up, and avoiding generic responses to build trust. | Reported higher trust when the coach remembered prior context and followed up on earlier goals, and described repetitive or static replies as robotic and less credible. |
| Structure and accountability | Saw structured goal setting and reminders as important for sustaining motivation and accountability. | Described progress tracking and check ins as helpful for staying on track and feeling more productive. |
| Reflective feedback | Valued ACT-informed reflection but cautioned against clinical jargon. | Found visual reflections and values prompts useful for noticing patterns, although some wanted more depth or challenge and sometimes ignored views that did not feel accurate. |
| Engagement over time | Recommended flexible engagement and easy re-entry rather than fixed daily routines. | Showed two broad trajectories: some built a habit around check ins, while others dropped off during busy periods and did not always notice opportunities to re-engage. |

## 6.2 Designing Conversational Support Around Follow-Through

Our findings point to a way of designing conversational well-being support that focuses on helping students follow through on their goals. Students rarely described *GROW* in terms of isolated emotional conversations. Instead, they consistently framed its usefulness in terms of follow-through: returning for check-ins, reminders, and visible progress over time. Even when interactions involved stress, frustration, or emotional venting, these moments were rarely treated as the endpoint. They were linked back to action, such as deciding next steps, breaking goals into smaller parts, or reaffirming commitments.

Designing around follow-through, in this sense, means folding reflection into concrete planning and adjustment around students' goals, rather than treating reflection as a separate emotional activity. Many conversational well-being systems foreground emotional reflection, mood tracking, symptom check-ins, journaling, or brief CBT-style exercises as the main unit of interaction [10, 24, 27, 60, 80]. Our findings add to this work by showing how students can experience support primarily through help with follow-through on values-aligned goals, with emotional talk playing a supporting role. This emphasis is also in line with goal-setting and self-regulation research that highlights ongoing planning, feedback, and monitoring rather than one-time goal declarations [51, 71].

## 6.3 Designing for Trust Through Continuity and Boundaries

Using a system repeatedly and incorporating it into everyday routines requires trust. Prior work on relational agents and digital mental health tools shows that trust and alliance can develop over repeated interactions with an automated system, and that these relationships can support adherence and long-term engagement [10, 13]. In our study, trust in *GROW* also built up over time, but it depended less on affective warmth in individual turns and more on whether the system behaved attentively and consistently within students' goal work.

Students described trusting the system when it demonstrated continuity: remembering prior context, following up on earlier goals, asking specific questions, and avoiding repetitive or generic responses. When trust broke down, they did not say that the coach felt insufficiently caring. Instead, they described it as robotic, shallow, or inattentive. In other words, trust failures were experienced as failures of follow-through and specificity rather than failures of empathy in any single message.

Many conversational well-being systems are designed around short, self-contained interactions that deliver particular cognitive or emotional techniques [24, 25, 35, 85]. In our findings, trust and engagement did not seem to hinge on a specific exercise or an emotionally resonant response. Instead, they accumulated across interactions, as students returned to *GROW* at moments when they needed support and found that it still reflected what they had been working toward over time. For students, this kind of continuity signaled reliability more strongly than expressive language alone, suggesting that designing for trust involves making the system's memory and follow-up visible.

Trust was also shaped by perceived boundaries. Experts emphasized the importance of role clarity and ethical positioning, and students reflected this concern in how they chose to disclose. Some felt comfortable sharing sensitive details because the system was non-human, while others withheld information despite confidentiality assurances. Students seemed to calibrate trust not only based on how attentive the system appeared, but also on whether its role and limits were clear.

Taken together, these findings suggest that trust in conversational well-being systems is built through a combination of consistent attentiveness over time and clearly communicated boundaries on what the system can and cannot offer, rather than through momentary expressions of empathy alone.

## 6.4 Designing Representations to Support Reflection in Conversation

In many conversational well-being systems, reflection is primarily supported through dialogue [24, 35]. In *GROW*, however, students often described reflective insight as emerging through dashboard representations rather than conversation alone. While the chat was the space where goals were expressed, updated, and negotiated, features such as goal progress indicators, values alignment and communication style summaries helped students notice patterns and reconsider priorities across time.

From a design perspective, this points to representations that act as cues for reflection rather than as diagnostic scores. Views of progress, check-in patterns, and values can be scanned quickly, while conversation provides nuance and context. This use of lightweight feedback is similar to how dashboards and self-monitoring apps present well-being information to support ongoing self-management rather than formal assessment [50, 78], and in *GROW* it allowed students to both talk through their experiences and revisit what they were working toward.

## 6.5 Integrating Well-Being Frameworks into Conversational Design

Students did not engage with *GROW* to learn ACT or SMART as frameworks. Instead, our findings suggest that these models shaped how interactions unfolded when students set goals, revised them, encountered obstacles, or reflected on values. In prior work, ACT interventions for students are often delivered as structured programs, with concepts and skills introduced explicitly across online or blended sessions, and SMART is commonly used to scaffold how goals are initially set in digital tools [33, 46, 67]. In *GROW*, these frameworks functioned more as design resources than as topics of instruction.

Across the findings (Sections 5.4.1–5.4.3), experts described ACT and SMART as giving the system structure and credibility, while cautioning against clinical language or jargon. Students did not name the frameworks, but they talked about values reminders, manageable next steps, and visible progress as helpful parts of the experience. These elements are in line with work on goal setting and self-regulation, where clear goals, feedback, and connection to underlying motives can support feelings of competence, focus, and purpose [2, 36].

In *GROW*, ACT and SMART therefore operated mainly as background structure rather than as content to be taught through conversation. They informed how questions were phrased, how larger goals were broken down, and how progress and values were represented, without overwhelming students with terminology or formal theory. Using well-being frameworks in this way lets designers draw on established psychological work to shape a goal-progressing workflow that combines conversational support, visual representations, and integration into everyday tools, while keeping the system positioned as a non-clinical support rather than a therapeutic program.

### 6.6 Design Implications for Goal-Oriented Conversational Well-Being Systems

Building on the discussion, our findings suggest the following design implications for goal-oriented conversational well-being systems:

- **Design for attentive follow-up.** Let the system remember ongoing goals and earlier context, and use that information in follow-up. Students should experience the coach as attentive and consistent across time, rather than as responding to each message in isolation.
- **Pair conversation with lightweight representations.** Use conversation as the place where goals and everyday stressors are worked through, and simple dashboard summaries as a way to see how things are going over time. Progress indicators, check-in patterns, and values views can provide quick overviews that sit alongside the nuance of chat.
- **Use representations as starting points.** Treat dashboards and other representations as approximate prompts that help students notice patterns and reconsider priorities, rather than as diagnostic assessments. Summaries should highlight what might be worth reflecting on, without claiming to fully capture a student's state.
- **Use well-being frameworks as background structure.** Let frameworks such as ACT and SMART guide how questions are phrased, how larger goals are broken down, and how progress and values are surfaced, without presenting these frameworks as material to be learned. This keeps interactions structured and grounded in theory while avoiding clinical jargon.
- **Be explicit about scope and limits.** Make the system's non-clinical role clear, including what it can and cannot offer and how data is used. Simple explanations of boundaries and capabilities can help students decide what to share and how to rely on the system.

### 6.7 Limitations and Future Work

While the study demonstrates promising early results, several limitations should be considered.

First, the deployment period was limited to one week. This scope enabled observation of how students initially integrated *GROW* into their routines, responded to conversational coaching, and engaged with reflective and progress-tracking features. The PERMA-Profiler was used to capture short-term changes in well-being within this window. Future work could extend this design across multiple weeks or academic cycles to examine how engagement patterns and well-being trajectories, including PERMA dimensions, evolve over time, particularly during periods of sustained academic pressure.

Second, the study did not include a control condition. As a result, changes in PERMA scores and students' reported experiences cannot be attributed solely to *GROW*, and may also reflect novelty or other aspects of the academic context during the study period. Future work could compare goal-oriented conversational support with alternative digital tools or standard support practices to better understand the specific contribution of systems like *GROW*.

Participants' experiences of conversational responsiveness also varied. While many described meaningful follow-up and contextual awareness, some encountered repetitive phrasing or more neutral responses. These differences reflect challenges in maintaining conversational depth, variation, and consistency in prompt-based large language model systems. Usage of the app decreased after the mandatory phase, limiting conclusions about longer-term use. Future iterations could explore more robust strategies and prompting mechanisms to better support sustained, context-sensitive interaction over longer periods, and to understand how variations in responsiveness affect trust, engagement, and perceived usefulness.

## 7 Conclusion

This work introduced *GROW*, a conversational AI system designed to promote student well-being through goal setting, reflection, and values alignment. Grounded in Acceptance and Commitment Therapy (ACT) and the SMART goal framework, *GROW* translates abstract intentions into concrete, values-based actions. Through a two-phase evaluation, practitioners confirmed the theoretical and contextual soundness of the system, while students demonstrated meaningful engagement with goal setting and values alignment, showing modest, directionally positive improvements across PERMA well-being domains. These findings suggest that structured digital interventions can complement university support systems by encouraging reflection, motivation, and incremental behavioral change.

### Acknowledgments

This work is supported by the NYUAD Center for Interdisciplinary Data Science & AI (CIDSAI), funded by Tamkeen under the NYUAD Research Institute Award CG016.

## A System Prompts

### A.1 Introduction Phase Prompt

**Prompt for the Introduction Phase**

**Task**: INTRODUCTION PHASE. Get to know the student naturally through conversation; do NOT set goals yet. REQUIRED SEQUENCE (ask conversationally, ONE at a time):

Basic Info (5):
- Name, college year, major
- How they feel (today + generally)
- Emotional awareness (high/medium/low in their words)
- Coping style (healthy/mixed/avoidant in their words)
- What encouragement helps (praise/progress/achievement/effort)

Personality (5) - ask naturally, like getting to know a friend:
- "Are you the type who gets excited about trying new things and exploring ideas?" (Open-mindedness: high/moderate/low)
- "How are you with staying organized and disciplined - natural for you or more of a struggle?" (Conscientiousness: high/moderate/low)

- "Would you describe yourself as more outgoing or more reserved?" (Extraversion: high/moderate/low)
- "When it comes to relationships, do you really focus on harmony and getting along with people?" (Agreeableness: high/moderate/low)
- "How often do you experience worry or stress - a lot or not so much?" (Neuroticism: high/moderate/low)

CONVERSATION STYLE: Make it feel natural - use their name, show genuine interest ("That makes sense", "I can see that"), make connections between answers ("So you're in CS and pretty organized - that probably serves you well"). Keep it friendly, strictly not clinical or diagnostic.

Light ACT touches only (breath check-in, normalize feelings, values preview).

Flow:

- You MUST collect all 10 pieces (5 basic + 5 personality) before proceeding.
- ONLY call `saveProfile` when you have complete demographic, ALL five `personality_traits`, and `mental_health_profile` data.
- After saving, say: "Thanks for sharing all of that with me. I feel like I'm getting a good sense of who you are. Now I'm curious about what really matters to you in different areas of your life."
- Do NOT include any phase markers yet. The system will automatically transition to the values check-in.

## A.2 Goal Setting Prompt

**Prompt for the Goal Setting Phase**

**Task**: GOAL SETTING PHASE. Start from values (ACT), then shape SMART goals.
You have this profile: *[user_profile]*
Recent conversation: *[history]*
Flow (ONE question at a time):

- Values cue: "What feels most important for your well-being right now?" (offer everyday examples only if needed, such as sleep, stress relief, friendships, or study rhythm).
- Pick one area to start.
- Build a SMART goal + define a weekly measure that enables auto progress:
  - Specific (what exactly?)
  - Measurable (use count, frequency, or minutes, with a clear weekly target).
  - Achievable (tiny first step)
  - Relevant (tied to their value)
  - Time-bound (clear timeframe)
- Plan 1–3 tiny actions for week one (committed action).
- Identify likely obstacles and choose a brief coping response using acceptance or defusion (for example, acknowledging the thought and continuing with a 2-minute starter action).
- Confirm the goal feels realistic.
- Ensure `measures` is phrased so progress can be inferred from chat (e.g., "3 wind-downs/week", "10 min mindfulness ×4/week").
- Call `saveProfile` once with JSON matching `ProfileJsonSchema`, including `mental_health_goals[]` objects with description, measures, timeframe, steps, obstacles, completed: false, progress: 0
- After the user indicates they have finished setting goals (typically 1–3 total), include the internal marker [ONGOING_PHASE] in your text to transition to the ongoing conversation phase.

## A.3 Active Coaching

**Prompt for the Active Coaching Phase**

**Task**: ONGOING CONVERSATION PHASE. Support progress with brief, practical tips.
Have profile + goals: *[user_profile]*
Recent conversation: *[history]*
Flow: If just came to ONGOING CONVERSATION PHASE from GOAL SETTING PHASE, ask if they are set and say that you'll be here when they need you and end conversation.

- Quick check-in ("How's today going?"). If the last 4 turns already covered whether they did a daily action (e.g., drank water), do not ask it again; instead acknowledge the answer and proceed.
- Take one goal at a time. Ask about progress/obstacles (short). Infer % progress from what they say
- When user confirms they want a snapshot ("yes", "sure", "show me"):
  - Simply show their current progress from the profile data above
  - Example: "Here's where you are: Your skincare wind-down goal is at 40% (you've done it 2 times this week, aiming for 5)."
- Offer one concrete next step (tiny, value-aligned).
- Use brief ACT-inspired micro-moves when helpful (for example: pausing to notice a feeling, acknowledging a thought without debating it, or gently refocusing on what matters).
- This is supportive, non-clinical coaching. Avoid diagnostic, therapeutic, or prescriptive mental health language.
- Celebrate a small win; normalize setbacks.
- If a goal is done, mark completed. If all done, move to [GOAL_SETTING_PHASE].
- End with a tiny, doable action before next time.
- Call `saveProfile` tool once per update turn when there is new or modified goal progress, following the same `ProfileJsonSchema`, and include the updated `mental_health_goals[]` object with description, progress, completed, lastUpdated.

> Skip the tool call if the user's response indicates no change in progress, to avoid redundant saves.
>
> Begin by greeting them and asking how they've been since last time. Keep it brief.

### A.4 Bull's Eye Value Survey

**Prompt for the Bull's Eye Value Survey Phase**

**Task**: VALUES CHECK-IN (BEVS). Map what matters across four life domains and capture 1–7 "closeness to values" scores, using warm, brief, conversational turns.

Recent conversation: *[history]*

Context:
- Current step: *[currentStep]*
- Current domain index: *[domainIndex]* (*Work/Studies, Relationships, Personal Growth/Health, Leisure*)

Flow:
- If step = intro: Briefly introduce the "values check-in" across a few life areas and ask to start with the current domain. Keep it friendly and conversational. Track the start time internally (not visible to user).
- If step = collect_values: Ask "In *<domain>*, what kind of person do you want to be or what matters to you?" Collect a short, natural answer (e.g., *learn deeply*, *show up for friends*, *care for health*). Advance to `currentStep: collect_scores`.
- If step = collect_scores: Ask "On a scale of 1–7, how close are your actions to your values in *<domain>*?" Validate responses within range. If domainIndex < 3, move to the next domain; otherwise, proceed to `confirm`.
- If step = confirm: Summarize in 2–3 lines, highlighting the lowest and highest domains. Suggest one tiny, values-aligned action (e.g., *send one supportive text*, *2-min wind-down*). Ask "Shall I save this values check-in and move on?"
- If user confirms (yes/done): Call tool `saveProfile` once with the complete BEVS object:

  `{ startedAt, completedAt, currentStep: "done", domainIndex: 3, domains[], assessments[] }`

- If step = done: Thank them, say "Now that I know what matters to you, let's set a goal that aligns with your values" and include `[GOAL_SETTING_PHASE]` to move forward.

**Tone & constraints:**
- Never say "BEVS" to the user. Always refer to it as "values check-in" or "values mapping."
- Keep the tone warm, conversational, and brief with one question per turn.
- Provide examples of values naturally to make it feel human, not survey-like.
- Never show JSON or tool details in user-visible text.

### A.5 Communication Style

**Prompt for the Communication Style**

You are analyzing a user's communication style for coaching personalization. ONLY consider lines that start with "User:" in the transcript provided. Ignore any "Coach:" lines.

Use the transcript as primary evidence and the derived metrics below as tie-breakers. Do not default to "casual", "short", "neutral", or "experience-based" unless the transcript clearly supports it.

Categories and guidance:
- **Tone**: "formal" (polite forms, complete sentences, minimal slang) vs "casual" (slang, contractions, informal phrasing).
- **Length**: "short" (typically ≤ 12 words per message) vs "long" (typically more than 12 words on average).
- **Emotional style**: "expressive" (emotion words, emojis, exclamations, self-disclosure) vs "neutral" (matter-of-fact, low affect).
- **Thinking style**: "data-driven" (numbers, structure, explicit reasoning, references) vs "experience-based" (anecdotes, personal examples, intuitive language).

Return a single compact JSON object with exactly these keys and values:

```
{"tone": "formal|casual", "length": "short|long",
 "emotional_style": "expressive|neutral",
 "thinking_style": "data-driven|experience-based"}
```

Output ONLY JSON (no prose).